\documentclass[]{revtex4}
\usepackage{epsfig}
\begin{document}

\title[Bogoliubov excitation spectrum in anharmonic traps]{Bogoliubov excitation spectrum in anharmonic traps}

\author{E Gershnabel, N Katz}
\author{ E Rowen}
\email{eitan.rowen@weizmann.ac.il}
\author{N Davidson}
\affiliation{Department of Physics of Complex Systems,\\
Weizmann Institute of Science, Rehovot 76100, Israel}

\begin{abstract}
We study the linearized Bogoliubov excitation spectrum of
infinitely long anharmonically trapped Bose-Einstein condensates,
with the aim of overcoming inhomogeneous broadening. We compare
the Bogoliubov spectrum of a harmonic trap with that of a
theoretical flat-bottom trap and find a dramatic reduction in the
inhomogeneous broadening of the lineshape of Bogoliubov
excitations. While the Bragg excitation spectrum for a condensate
in a harmonic trap supports a number of radial modes, the flat
trap is found to significantly support just one mode. We also
study the excitation spectrum of realistic anharmonic traps with
potentials of finite power dependence on the radial coordinate. We
observe a correlation between the number of radial modes and the
number of bound states in the effective potential of the
quasi-particles. Finally we compare a full numerical
Gross-Pitaevskii simulation of a finite-length condensate to our
model of infinite, linearized Gross-Pitaevskii excitations. We
conclude that our model captures the essential physics.
 \end{abstract}

\maketitle

\section{Introduction}
\label{introduction}

The spectrum of weak excitations over the ground state of a
Bose-Einstein condensate (BEC) has been found to obey the
Bogoliubov dispersion relation \cite{ours}. There is however, a
finite width to the  dynamical structure factor $S(k,\omega)$
which characterizes the response of the BEC to an exciting field
with momentum $k$ and frequency $\omega$. Usually the exciting
field is in a Bragg configuration \cite{Ketterle_Doppler}, where
$k$ is the momentum difference of two laser beams, and $\omega$ is
the detuning between the two.  There is a small intrinsic width
originating from the Beliav decay of quasi-particles which is
usually dominated by broadening due to finite time of the Bragg
pulse and inhomogeneity of the condensate. The latter heavily
depends upon the trap geometry. In some cases the wavelength of an
excitation is much shorter than the condensate dimensions. Under
these circumstances  it is justified to employ the local density
approximation (LDA) \cite{LDA}. For cigar shaped condensates, the
radial dimension is sometimes comparable to the excitation
wavelength, leading to the breakdown of the LDA, and the
appearance of radial modes \cite{radial_modes,tozzo}. Avoiding
inhomogeneous broadening is possible in the spectral domain using
echo spectroscopy where a degenerate Bragg pulse transfers
Bogoliubov excitations with wavenumber $+k$ to $-k$ \cite{Erez},
and in the time domain where rapid oscillations cause a
suppression of inhomogeneous mean field and of Doppler dephasing
\cite{Nadav_reduction}. Suppression of the inhomogeneity
mechanisms, allows longer coherence times, and opens the
possibility of studying the homogeneous broadening mechanisms,
that reflect the intrinsic decoherence processes of the bulk
excitations, e.g. elastic collisions with the BEC \cite{Collis}.

In this paper we propose the use of a flat-bottom trap in order to
reduce the inhomogeneous broadening of  Bogoliubov excitations in
a BEC. We restrict ourselves to the cigar shaped geometry, in
which the inhomogeneity is mainly manifested by the presence of
many radial modes. By linearizing the Gross-Pitaevskii equation
(GPE), assuming a uniform and infinite potential in the axial
direction and using a quasiparticle projection method
\cite{tozzo}, we obtain the Bogoliubov excitation spectrum for a
flat-bottom trap. Then, we compare the above results to the
excitation spectrum of a harmonically trapped BEC. A clear
reduction in the inhomogeneous broadening is evident by comparing
between the two spectra. This result, although intuitive, is not
trivial for two reasons. The first is that even in a flat trap the
density is not constant, but varies over a distance characterized
by the healing length $\xi$.

The second reason is that even a hypothetical homogeneous finite
condensate supports several radial modes. Indeed, we find that it
is the gradual decay of the density to zero at the trap boundary
that reduces the coupling to the high order radial modes. The
envelope of both harmonic and flat trap spectra are found to match
the LDA \cite{LDA}. We then examine the excitation spectrum of a
condensate trapped in the power-law potential
$V=\kappa_\rho\rho^p$, where $\rho$ is the radial coordinate, and
$p$ is an even integer. We observe the substantial excitation of a
second radial mode in the spectrum, and demonstrate its
suppression using higher order potentials. The transition between
potentials supporting one or two radial modes is explained by an
effective potential picture. Finally, we solve the full GPE and
find the Bogoliubov spectrum for a realistic, experimentally
feasible \cite{flat} three-dimensional axially elongated trap,
confined both in the radial and axial direction using a high-order
potential. The spectrum obtained by the linearized approach where
the potential is axially uniform, and by the full GPE for an
elongated condensate, are found to be qualitatively similar due to
the tight radial confinement.

The outline of the paper is as follows: in section
\ref{cylindrical}, we study the ground state and the
quasiparticles of the flat-bottom trap, then in section
\ref{bogoliubov}, a comparison between the flat-bottom trap and
the harmonic trap excitation spectra is given. In section \ref{nonHarmonic}
we study high order anharmonic potentials, still axially infinite,
and section \ref{axial} is devoted to the full GPE calculation of
a three dimensional elongated anharmonic trap.

\section{Ground state and Bogoliubov quasi-particles in an infinite cylinder}
\label{cylindrical}

Since the traps discussed in this paper are cigar shaped, the
dominant contribution to the inhomogeneity originates from the
trap geometry in the radial direction. Thus we first assume an
infinite trap in the axial direction $\hat{z}$.  We solve the
linearized GPE for a flat-bottom trap, uniform in the axial
direction. In order to compare with our harmonic trap, we fix the
chemical potential $\mu/h=2.06$ kHz (in the Thomas-Fermi
approximation) and the number of atoms per unit length $N=10^5$
atoms per $L=52.2\mu$m to coincide with our BEC experimental
parameters \cite{ours}.

Our flat-bottom trap potential, is a cylindrically symmetric
hard-wall confinement in the radial direction:
\begin{equation}
V_T(\mathbf{r})=\left\{
\begin{array}{c}
0 \\
\infty
\end{array}
\right.
\begin{array}{c}
\rho<R_{TF} \\
otherwise
\end{array}
\
\end{equation}
where $R_{TF}$ is the radial Thomas-Fermi radius, which is
determined by the constraints on $\mu$ and $N/L$. We calculate the
Bogoliubov spectrum for the flat-bottom trap following the
derivation of \cite{tozzo} for harmonic traps. Briefly, in the
linear regime, the condensate wavefunction can be expanded as:
\begin{equation}
\Psi(\mathbf{r},t)=e^{i(\mu/\hbar)t}\{\psi_{0}(\mathbf{r})+\sum_{j}[c_{j}u_{j}(\mathbf{r})e^{-i\omega_{j}
t }+c_{j}^{*}v_{j}^{*}(\mathbf{r})e^{i\omega_{j} t }]\}
\label{linear_wave}
\end{equation}
where the $c_{j}$'s are constants, and $u$ and $v$ are the
quasi-particle amplitudes, that
obey the following orthogonality and symmetry relations:
\begin{equation}
\int d \mathbf{r}\{ u_{i}u_{j}^{*}-v_{i}v_{j}^{*}\}=\delta_{ij};
\space \int d \mathbf{r}\{
u_{i}v_{j}-v_{i}u_{j}\}=0\label{relations}
\end{equation}
The wavefunction expansion in equation \ref{linear_wave} is
inserted into the GPE: $i\hbar \frac{\partial}{\partial
t}\Psi=\left(-\frac{\hbar^2\nabla^2}{2m}+V_T+g|\Psi|^2\right)\Psi$,
where $g=\frac{4\pi \hbar^2 a}{m}$ is the coupling constant, $m$
is the atomic mass and $a$ is the $s$-wave scattering length. For
a cylindrical condensate, infinite in the axial direction, we
define $\psi_{0}(\rho,z)=\phi_{0}(\rho)/L$,
$(u,v)_{n,k}(\rho,z)=\frac{1}{\sqrt{L}}e^{ikz}(u,v)_{n,k}(\rho)$
as the new ground state wavefunction and the quasi-particle
amplitudes, respectively. Gathering terms to zero order $u,v$
yields  the stationary GPE:
\begin{equation}
\left(-\frac{\hbar^{2}\nabla^{2}_{\rho}}{2m}+V_{T}(\rho)+\frac{gN}{L}\phi_{0}(\rho)^{2}\right)\phi_{0}(\rho)=\mu
\phi_{0}(\rho)
 \label{Stationary}
\end{equation}
The next order gives the coupled equations:
\begin{eqnarray}
\hbar\omega_{n,k} u_{n,k}(\rho)=
[-\frac{\hbar^{2}\nabla_{\rho}^{2}}{2m}+\frac{\hbar^2 k^2
}{2m}+V_T(\rho)-\mu+\frac{2gN}{L}\phi_{0}^{2}(\rho)]u_{n,k}(\rho)
+\frac{gN}{L}\phi_{0}^{2}(\rho)v_{n,k}(\rho) \label{linear1}
\end{eqnarray}
\begin{eqnarray}
-\hbar\omega_{n,k} v_{n,k}(\rho)=
[-\frac{\hbar^{2}\nabla_{\rho}^{2}}{2m}+\frac{\hbar^2 k^2
}{2m}+V_T(\rho)-\mu+\frac{2gN}{L}\phi_{0}^{2}(\rho)]v_{n,k}(\rho)
+\frac{gN}{L}\phi_{0}^{2}(\rho)u_{n,k}(\rho) \label{linear2}
\end{eqnarray}
also known as the linearized GPE or
Bogoliubov equations.

\begin{figure}
\begin{center}
\epsfbox{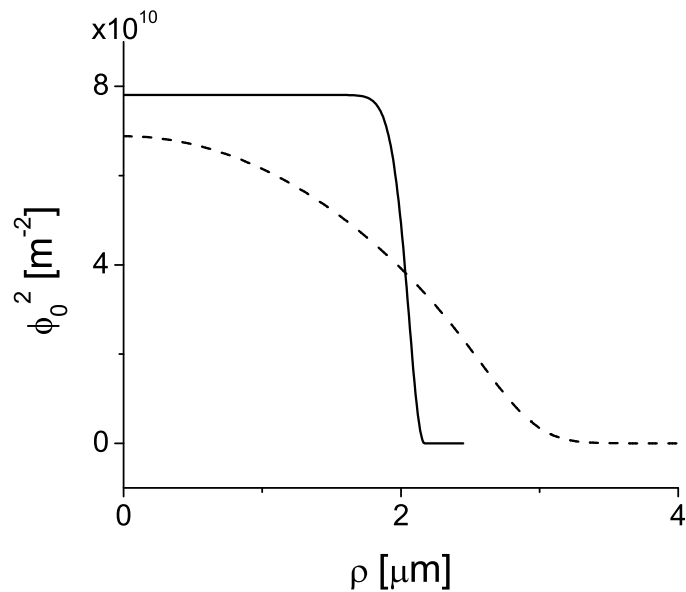}
\end{center}
\caption{\label{wave_func} Ground state radial wavefunctions of
axially infinite harmonic (dashed line) and flat-bottom (solid
line) traps obtained by solving the stationary GPE with imaginary
time evolution. }
\end{figure}

First, we find the ground state wavefunction from equation
\ref{Stationary} by applying imaginary time evolution
\cite{imaginary} to the Thomas-Fermi wavefunction
\cite{Dalfovo_review}. Figure \ref{wave_func} shows the ground
state wavefunctions for the flat-bottom trap (solid line) and for
an axially infinite harmonic trap (dashed line), with a confining
potential of $V_T=\frac{1}{2}m\omega_{\rho}^2\rho^2$ where
$\omega_{\rho}=2\pi\times 226 $Hz \cite{Erez}. Indeed, the
harmonic trap wavefunction is  similar to the Thomas Fermi ansatz,
and the flat trap has a ground state in which the density drops to
zero at $R_{TF}$ over a length scale $\xi=(8\pi n(0)a)^{-1/2}$
\cite{Dalfovo_review}, where $n(0)$ is the peak density.

In order to find the Bogoliubov frequencies $w_{n,k}$ and the
quasi-particles amplitudes $(u,v)_{n,k}$, the relaxed wavefunction
is substituted in equations \ref{linear1} and \ref{linear2}.  The
Bogoliubov equations \ref{linear1} and \ref{linear2} are an
eigenvalue problem and are solved by expanding the functions
$u(\rho)$ and $v(\rho)$ using a complete set of zero order Bessel
functions. The problem is thus reduced to a matrix diagonalization
and is numerically solved \cite{tozzo}. We take the excitation
momentum $k$ to be  $k=8.06\mu m^{-1}$ as in \cite{Erez}.

\begin{figure}
\begin{center}
\epsfbox{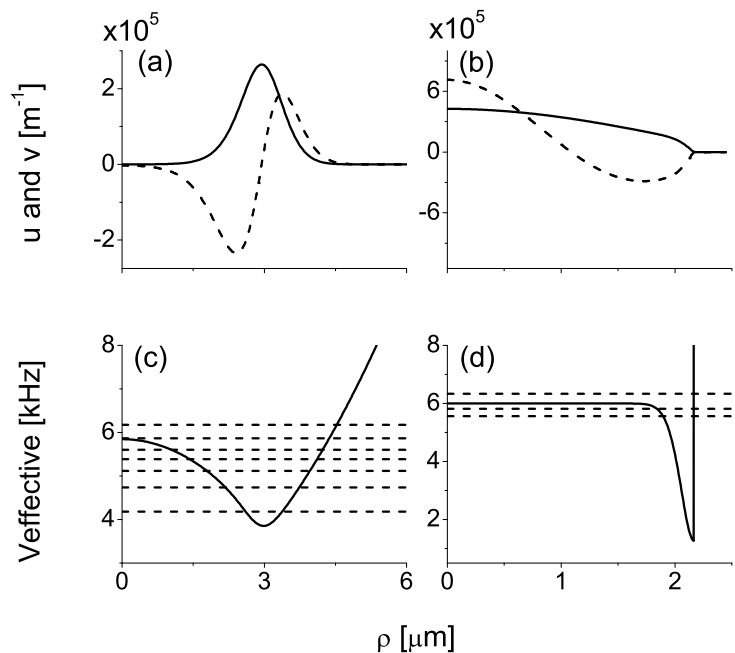}
\end{center}
\caption{\label{quasi_particles_wave} Quasi-particles amplitudes
$u_{n,k}$ and effective potentials. (a) Harmonic trap, $u_{0,k}$
(solid line) and $u_{1,k}$ (dashed line). (b) Flat-bottom trap,
$u_{0,k}$ (solid line) and $u_{1,k}$ (dashed line). (c) Harmonic
trap, effective potential (solid line) and Bogoliubov frequencies
(dashed line) . (d) Flat-bottom trap, effective potential (solid
line) and Bogoliubov frequencies (dashed line).}
\end{figure}

The calculated lowest order quasi-particle amplitudes $u_{0,k}$
(zero order, solid lines) and $u_{1,k}$ (first order, dashed
lines) are shown in figure \ref{quasi_particles_wave}(a) and (b),
for the harmonic and flat-bottom trap, respectively. For the
$n^{th}$ order radial modes, both $u$ and $v$  have $n$ nodes.

Comparing figures \ref{quasi_particles_wave}(a) and (b) reveals a
dramatic difference. For the flat bottom trap (figure
\ref{quasi_particles_wave}(b)) $u_{0,k}$ and $u_{1,k}$ are mainly
centered around $\rho=0$ whereas for the harmonic trap (figure
\ref{quasi_particles_wave}(a)) they are significantly shifted.

To qualitatively explain this effect we approximate equation
\ref{linear1} as:

\begin{equation}
\hbar\omega_{n,k} u_{n,k}(\rho)=
[-\frac{\hbar^{2}\nabla_{\rho}^{2}}{2m}+\frac{\hbar^2 k^2
}{2m}+V_T(\rho)-\mu+\frac{2gN}{L}\phi_{0}^{2}(\rho)]u_{n,k}(\rho)
\label{free_approximate}
\end{equation}
where we neglect the contribution of $v$, which are small for the
$k\xi>1$ regime considered here. The two Bogoliubov equations
\ref{linear1} and \ref{linear2} are thus reduced to a
 Schr\"{o}dinger equation for the excitations, with the
effective potential $V_{eff}=\frac{\hbar^2 k^2
}{2m}+V_T(\rho)-\mu+\frac{2gN}{L}\phi_{0}^{2}(\rho)$
\cite{stringaribook}. In figure \ref{quasi_particles_wave}(c) and
(d) we plot the effective potential for the harmonic and
flat-bottom trap, respectively. The dashed lines are the lowest
eigenvalues $\omega_{n,k}$ of equations \ref{linear1} and
\ref{linear2}, without neglecting $v$. The effective potential of
the harmonic trap can be approximated for the lowest modes as a
harmonic potential, leading to $u(\rho)_{n,k}$ which resemble
eigenstates of the harmonic trap for $\omega_n$ sufficiently deep
in the effective trap. However, for eigen-energies
$\hbar\omega_{n,k}$ which approach the effective trap height,
there is a nonvanishing amplitude that interferes constructively
near $\rho=0$ due to radial symmetry, so that the amplitude is not
concentrated in the well. This happens even for the lowest mode in
the flat trap. The well in this case is extremely narrow,
originating only from the spatially smoothing of the $\phi_0$ at
the walls, supporting no ``bound states'' concentrated in the
well.

\section{Bogoliubov Bragg spectroscopy of a flat-bottom trap}
\label{bogoliubov}

Once the quasi-particles amplitudes and the Bogoliubov frequencies
are obtained, they can be used to derive the Bogoliubov Bragg
spectrum, using the quasi-particle projection method \cite{tozzo}.
Briefly, in this method  the $c_j$ coefficients in equation
\ref{linear_wave} are taken to be time-dependent and then
substituted  into the GPE, which includes now also the Bragg
external potential $V_B\cos(kz-\omega t)$, where $V_B$ is the
strength of the Bragg potential. The momentum given to the
condensate, in the Bragg process, defined as:
\begin{equation}
\label{first} P_{z}(t)=\frac{\hbar}{2i}\int d\mathbf{r}
\Psi^{*}(\mathbf{r},t)\frac{\partial}{\partial
z}\Psi(\mathbf{r},t)+c.c.
\end{equation}
is found to be \cite{tozzo}:
\begin{eqnarray}
P_{z}(t)=\frac{NqV_{B}^{2}t^{2}}{4\hbar}\sum_{n}|W_{n,k}|^{2}
[(\frac{sin[(\omega_{n,k}-\omega)t/2]}{(\omega_{n,k}-\omega)t/2})^{2}-(\frac{sin[(\omega_{n,k}+\omega)t/2]}{(\omega_{n,k}+\omega)t/2})^{2}]
\label{second}
\end{eqnarray}
 where $W_{n,k}=2\pi\int d\rho \rho
[u^*_{n,k}(\rho)+v^*_{n,k}(\rho)]\phi_0(\rho)$. $W_{n,k}$ can be
considered as a weight function that determines
the weight of each radial mode.
This  weight is given by the overlap
between the ground state and the quasi-particles amplitudes
$(u,v)_{n,k}$.
\begin{figure}
\begin{center}
\epsfbox{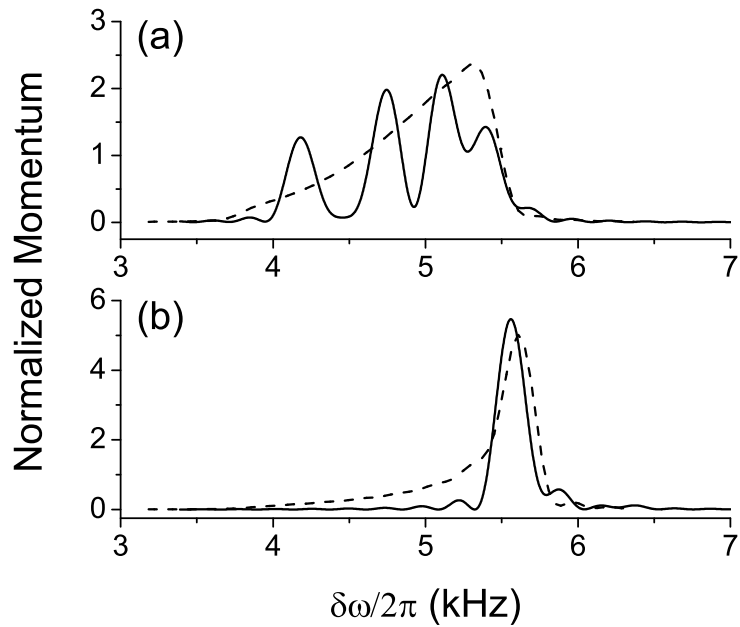}
\end{center}
\caption{\label{linearGraphs} Bragg spectra for Bogoliubov
(extremely weak Bragg pulses) excitations. (a) Harmonic trap. (b)
Flat-bottom trap. Solid line corresponds with the linearized GPE
solution, dashed line corresponds with the LDA. The $y$ axis is
the quantity $P_{z}/(NV_B^2\hbar k)$ in units of
$(\hbar\omega_{\rho})^{-2}$.}
\end{figure}

Figure \ref{linearGraphs}a and \ref{linearGraphs}b show the
linearized spectrum (solid lines) obtained from equation
\ref{second} for the $k=8.06\mu m^{-1}$ momentum, after
application of a $4.2$ ms Bragg pulse, in the harmonic and
flat-bottom trap, respectively. The harmonic trap spectrum of
figure \ref{linearGraphs}(a) shows four resolved radial modes. It
is similar to the Bragg spectrum obtained from full GPE
calculation, for harmonically trapped elongated BEC (figure 2a in
\cite{Erez}), as also noted in \cite{tozzo}. The lowest radial
mode has a small overlap with the ground state, due to its ring
shape, as seen in figure \ref{quasi_particles_wave}(a). This ring
shape can be explained using energy considerations. Due to
exchange symmetry, the mean-field energy of an excitation is twice
that of the ground state,
 making it more favorable to concentrate the
quasiparticle in a ring around the ground-state. In the effective
potential picture, the minimum of the well is near the
Thomas-Fermi radius for exactly this reason. The spectrum may thus
be qualitatively described by the following argument. For the
lowest radial mode, the overlap of the excitation wavefunction,
which is approximately gaussian, with the condensate wavefunction
is small. As the radial number $n$ increases, but remains small
enough for the effective potential to be approximated by a
harmonic oscillator the overlap with the ground state increases
since the quasiparticle wavefunction is wider. Upon further
increasing $n$, the oscillations in $u$ and $v$ wash out any
overlap with the condensate.

From figure \ref{linearGraphs}(b) it is apparent that the
flat-bottom trap supports mainly the first radial mode, strongly
suppressing the excitation of higher order modes, even though they
are valid solutions to the Bogoliubov equations. This suppression
is due to the small overlap $W_{n,k}$ between the quasiparticle
amplitude for $n>0$ and the ground state, as seen from in figures
\ref{wave_func} (solid line) and \ref{quasi_particles_wave}(b).
There are no states deep in the well of the effective potential,
and since only $u_{0,k}$ is nodeless it has the dominant overlap
with $\phi_0$. In the Thomas-Fermi limit of an effective potential
of a cylindrical well with infinite walls, the quasi-particle
states are Bessel functions  of order $0$, $J_0$. This
approximation yields a significant  overlap between $u$ and
$\phi_0$ for a number of radial modes. The overlap decays as $1/n$
for large $n$. Only when considering the relaxed wavefunction,
which is closer in shape to the function $J_0$, does the lowest
radial mode become dominant.

It is insightful to compare these results to the inhomogeneous
line shape calculated within the LDA \cite{LDA}, shown as dashed
lines in figure \ref{linearGraphs}. The LDA lineshapes were
obtained by integration over the structure factor weighted by the
inhomogeneous density
distribution of the relaxed wavefunctions. The
simplified LDA picture, whose basic assumptions are invalid here
indeed fails to produce the radial mode structure. Surprisingly,
it is seen from  figure \ref{linearGraphs} that it describes  the envelope of
the spectrum for both traps quite well, suppressing modes outside it
\cite{tozzo}.

\section{Other cylindrical anharmonic traps}
\label{nonHarmonic}

So far, an idealized infinite cylindrical trap was studied.
However, such a trap cannot be realized and therefore we consider
experimentally achievable   traps of finite high-order potentials
\cite{flat}. Here we show that even rather steep potential may
yield deviations from the single-mode spectrum of the flat-bottom
trap. This is done by evaluation of the Bogoliubov excitation
spectrum for $k=8.06 \mu m^{-1}$, in axially infinite traps with
anharmonic radial potentials of the form
$V_T(\mathbf{r})=\kappa_{\rho}\rho^p$ with $2\leq p \leq 20$,
using the procedure described above to solve the linearized GPE.
The number of atoms per unit length is chosen to be the same as in
section \ref{cylindrical}. The chemical potential is the same for
all traps, but different than that in section \ref{cylindrical}.

\begin{figure}
\begin{center}
\epsfbox{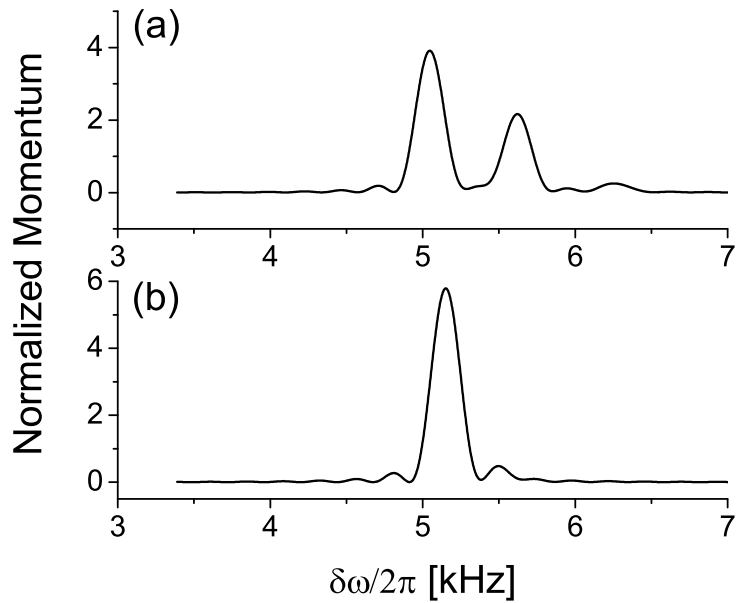}
\end{center}
\caption{\label{r10_r20} Bragg spectra for Bogoliubov excitations.
(a) $\rho^{10}$ trap. (b) $\rho^{20}$ trap. The $y$ axis is the quantity
$P_{z}/(NV_B^2\hbar k)$ in units of $(\hbar\omega_{\rho})^{-2}$.}
\end{figure}

\begin{figure}
\begin{center}
\epsfbox{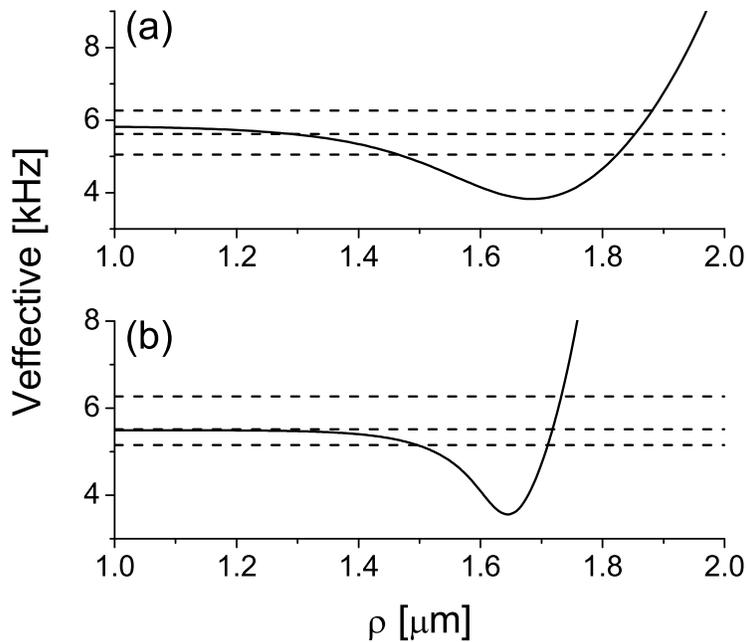}
\end{center}
\caption{\label{r20_r10_Veffect} Effective potentials (solid
lines) and linearized GPE Bogoliubov frequencies (dashed lines)
(a) $\rho^{10}$ trap. (b) $\rho^{20}$ trap.}
\end{figure}

The resulting spectra for the potential of $\rho^{10}$ and
$\rho^{20}$ are shown in figures \ref{r10_r20}(a) and (b). A
strong suppression of modes is apparent, in comparison to the
number of modes in the harmonic trap (figure
\ref{linearGraphs}(a)). The appearance of a second radial mode in
the $\rho^{10}$ and not in the $\rho^{20}$ trap seems surprising
considering the flatness of both potentials. The qualitative
difference between the two traps, which leads to such different
response can be seen in the effective potentials plotted in figure
\ref{r20_r10_Veffect}.
In the effective potential picture of the $\rho^{10}$ trap (figure
\ref{r20_r10_Veffect}(a)) we discern two trapped energy states,
whereas for the $\rho^{20}$ trap (figure \ref{r20_r10_Veffect}(b))
we observe only one ``bound'' eigen-state, while the higher modes
are now ``free''. This leads to a dramatic change in the Bragg
coupling matrix elements, suppressing all higher order radial
modes, although these modes are within the allowed LDA envelope.
Note that the this picture is only approximate due to the neglect
of the $v$ terms. Interesting corrections are expected at lower
$k\xi$, where the $v$ contribution is no longer negligible. Also,
the effective Schr\"{o}dinger equation is radial, leading to
constructive interference near $\rho=0$.

We calculated the Bragg spectra for the Bogolibuov excitations for
other values of $p$, and found a monotonic reduction in the weight
of the higher order radial modes with increasing steepness of the
radial potential.

\section{Realistic three-dimensional anharmonic traps}
\label{axial}

Finally, we compare between the spectra obtained above with the
linearized GPE for cylindrical traps, with no axial confinement,
and the spectra obtained for realistic elongated condensates
confined in similar experimentally-realizable three-dimensional
anharmonic traps. We consider one specific optical trap generating
a  potential
$V_T(\mathbf{r})=\kappa_{\rho}\rho^{20}+\kappa_{z}z^{12}$ which is
very similar to the one used experimentally for the atom-optics
billiard of \cite{friedman}.

For this we numerically solve the full GPE $i\hbar \partial _{t}
\psi=\{-\hbar^{2}\nabla^{2}/(2m)+V+g|\psi|^{2}\}\psi$, with
$V(\mathbf{r},t)=V_T(\mathbf{r})+V_{B}cos(kz-\omega t)$. We evolve
$\psi$ on a two dimensional grid (using the cylindrical symmetry
of the problem) $N_{z}\times N_{r}=2048\times 32$, using the
Crank-Nicholson differencing method. We ignore the influence of
gravity to keep the cylindrical symmetry of the problem.
Experimentally, gravitation may be compensated for using a
magnetic field \cite{wieman}.

\begin{figure}
\begin{center}
\epsfbox{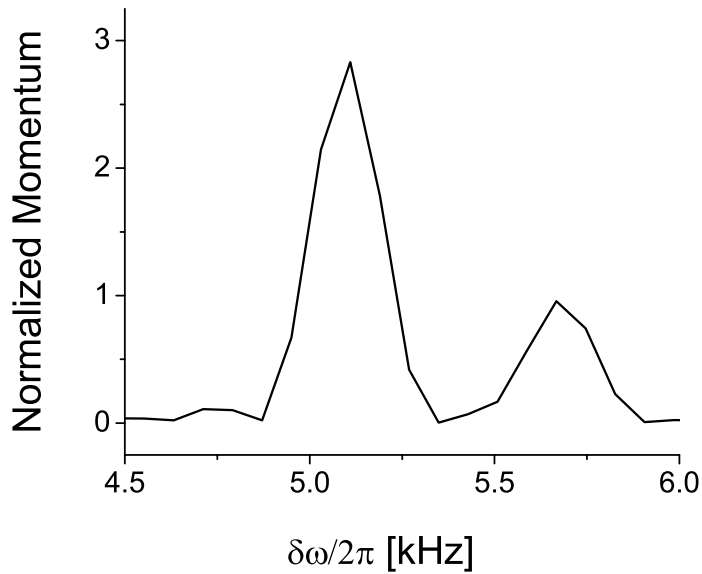}
\end{center}
\caption{\label{r20GPE} Bogoliubov spectra for the potential
$V_{T}(\mathbf{r})=\kappa_{\rho}\rho^{20}+\kappa_{z}z^{12}$, using
the full GPE.}
\end{figure}

The Bogoliubov Bragg spectrum thus calculated is shown in figure
\ref{r20GPE}.
Comparison between figure \ref{r20GPE} and figure \ref{r10_r20}b,
indicates a qualitative similarity between the spectra of the
``real'' three-dimensional traps and that of their idealized
axially infinite (two-dimensional) version, even though figure
\ref{r20GPE} includes additional inhomogeneity due to axial
confinement, as well as possible finite-size (Doppler) effects
\cite{Ketterle_Doppler,Nadav_reduction}. Thus we see the essence
of the inhomogeneous broadening is indeed captured by the radial
dependence of the trap.

\section{Conclusion}

In conclusion, we propose and study a possible approach to reduce
the inhomogeneity of trapped condensates by changing the
functional form of the trap. We show a reduction of the
inhomogeneous line-shape and suppression of the multiple peaked
structure in the Bragg Bogoliubov spectra by using anharmonic
traps with steep walls. The suppression is found to be monotonic
with the steepness of trap potential walls.  Linearized
calculations for simplified two-dimensional cylindrical traps give
qualitative understanding of the suppression of high-order radial
modes, due to decreasing overlap with the ground-state wave
function. Further LDA calculations we performed, suggest a
dramatic broadening of the spectra in the presence
 of gravity, emphasizing the necessity to compensate
for gravity in experimental realization of the anharmonic traps. Finally, full GPE
simulation of the spectra with previously used optical traps
confirm the experimental feasibility of this approach. The
reduction in inhomogeneous broadening  opens the possibility to
study the more intrinsic homogenous broadening mechanisms.
Experimental realization of BEC confined by anharmonic traps
should also yield new experimental opportunities, such as the
sharpening of the phase transition from superfluid to
Mott-insulator \cite{bloch} and rich variety of vortex phases \cite{vortex}.

 This work was supported in part by the Israel Ministry of
Science and the Israel Science Foundation. The authors wish to
express their gratitude to C. Tozzo and F. Dalfovo for their
assistance.

\section*{References}
%\Bibliography{<12>}

\end{document}